# Magneto-thermal-switching properties of superconducting Nb

Miku Yoshida[1], Md. Riad Kasem[1], Aichi Yamashita[1], Ken-ichi Uchida[2], and Yoshikazu Mizuguchi[1]*

[1]*Department of Physics, Tokyo Metropolitan University, Hachioji, Tokyo 192-0397, Japan.*
[2]*National Institute for Materials Science, Tsukuba, Ibaraki 305-0047, Japan*

E-mail: mizugu@tmu.ac.jp

Recently, thermal switching has been extensively studied because it is a key component for thermal management in electronic devices. Here, we show a huge magneto-thermal-switching ratio (MTSR) in pure Nb at temperatures lower than its superconducting transition temperature ($T_c$ = 9.2 K). The MTSR increases with decreasing temperature, and MTSR of 650% was observed at $T$ = 2.5 K under $H$ = 4.0 kOe. The thermal switching in superconductors with the huge MTSR will be useful for improvement of the performance of low-temperature electronic devices.

Thermal management technologies have been extensively studied due to recent development of nanoscale electronic devices.[1–3] The target phenomenon of this study is thermal switching, which can contribute to thermal management through active control of thermal conductivity ($\kappa$).[4–10] To achieve thermal switching characteristics, metal-insulator transition,[4,5] electrochemical intercalation,[6,7] electric field control of domain structures[8] and magnons[9], and magneto-thermal resistance effects[10,11] have been typically utilized. Among them, the large magnetic-field-induced $\kappa$ change of ~25 W/mK and the $\kappa$ change ratio of ~150% at room temperature was observed in a Co/CoFe spintronic multilayer film[10]. Although major investigations on thermal switching have aimed to be used at room temperature or higher temperatures, low-temperature thermal switching devices are also required due to rapid development of low-temperature electronic devices including superconducting quantum computers.[12–14] In this letter, we propose that superconductors are potential materials for low-temperature thermal switching because of a huge magneto-thermal-switching ratio (MTSR).

Generally, superconductors possess low $\kappa$ because of the suppression of carrier contribution of $\kappa$ ($\kappa_{el}$) below their superconducting transition temperature ($T_c$). One of the examples of application of the phenomenon is a high-$T_c$ superconductor current lead used for suppressing heat transfer into a superconducting magnet.[15,16] In addition, low-temperature $\kappa$ has been used in the field of physics of superconductors because precise evaluation of $\kappa$ under magnetic fields gives information for pairing mechanisms.[17–20] Actually, in the superconducting states, $\kappa_{el}$ is generally suppressed with decreasing temperature, and sensitivity of $\kappa_{el}$ to magnetic fields ($H$) has been detected. However, the magneto-thermal transport phenomena in superconducting states have not been discussed from the viewpoint of magneto-thermal switching (MTS). Therefore, we choose niobium (Nb) as a prototype superconductor and have studied its thermal conductivity under magnetic fields up to 4.0 kOe where the $\kappa$ became comparable to that in non-superconducting states. Thermal conductivity of Nb has been studied in the field of condensed matter physics[21,22] and superconductivity application for superconducting radio frequency cavities.[23,24] According to previous studies, $\kappa$ of Nb at low temperatures is dominated by $\kappa_{el}$, which is due to small contribution of phonon to $\kappa$; hence, Nb is suitable for evaluating the MTS properties in superconducting and non-superconducting states. This is different from the case of superconducting Pb, in which $\kappa_{el} > 50$ W/mK remains in superconducting states ($T = 2.5$ K).[25]

We measured $\kappa$ of a Nb sheet (99.9% purity, Nilaco) with a thickness of 0.1 mm using

Thermal Transport Option (TTO) of Physical Property Measurement System (PPMS, Quantum Design). The TTO works correctly under magnetic fields as demonstrated on Pb in Ref. 25. Four probes were fabricated using thin brass bars and Ag pastes on a Nb sheet (14.5 mm × 2.0 mm × 0.1 mm). The distance between two thermometer probes was 7.5 mm. Magnetic fields ($H$) up to 4.0 kOe were applied along the direction perpendicular to the generated thermal difference. The typical measurement period below 10 K was 30 seconds. Temperature dependence of electrical resistivity was measured by a four-probe method with an applied DC current of 5 mA on the PPMS. Magnetization was measured by a superconducting interreference device on Magnetic Property Measurement System (MPMS 3, Quantum Design).

Figure 1 shows the temperature dependence of $\kappa$ of the Nb sheet at various magnetic fields. At $H$ = 0 kOe, $\kappa$ begins to decrease at $T$ ~ 8.7 K, which is due to the emergence of superconducting states. With increasing magnetic field, the transition temperature, at which $\kappa$ exhibits a clear decrease, shifts to lower temperatures. At $H$ = 4.0 kOe, the trend of the $\kappa$ change due to the superconducting states is suppressed. The reproducibility of the $\kappa$-$T$ data was confirmed by other measurements on a Nb sheet (sample #2) as shown in Fig. S3.

To quantitatively analyze the MTS properties of Nb in the superconducting states, $\kappa$ at various temperatures were estimated and plotted in Fig. 2(a) as a function of $H$. It is clearly seen that the $\kappa(H)$ becomes flat at higher temperatures, which is due to the suppression of superconducting states with weak magnetic fields. Figure 2(b) shows the MTSR estimated by MTSR = [$\kappa(H)$ - $\kappa(0$ kOe$)$]/ $\kappa(0$ kOe$)$, and percentage is used for the vertical axis of the plot. At $T \leq 5.0$ K and $H \geq 2.0$ kOe, MTSR exceeding 50% was observed. With decreasing temperature, the MTS effect becomes huge, and MTSR of 650% was observed at $T$ = 2.5 K and $H$ = 4.0 kOe. Although one can easily find that lower temperatures and higher magnetic fields are preferred for a larger MTSR in Nb, we try to further discuss the evolution of MTS effects in Nb with superconductivity phase diagram. As shown in Fig. S4, the $\kappa$-$T$ loop does not show a clear hysteresis.

On the basis of electrical resistivity measurements under magnetic fields (see Fig. S1) and magnetization measurements (see Fig. S2), the magnetic field-temperature phase diagram was produced. In Fig. 3, upper critical fields ($H_{c2}$) and irreversibility fields ($H_{irr}$) estimated from the onset and zero-resistivity temperatures, respectively, were plotted. It is clear that the large MTSR exceeding 300% was observed at lower temperatures (lower than the transport $T_c$) only. On the other hand, moderate MTSR lower than 200% was observed in a wide $H$-$T$ range. The fact suggests that the moderate performance of MTS can be expected at flexible $H$-$T$ conditions in superconducting materials. We note that the lower critical fields ($H_{c1}$) are lower than 1.0 kOe for Nb as shown in Fig. S2(d). Therefore, the huge MTS effects are basically observed in the mixed states of superconductivity. To further develop MTS technique based on superconductors, elucidation of the key factors for a enhancing the MTSR and investigation on high-$T_c$ or exotic superconductors are desired.

In conclusion, we investigated the MTS characteristics of superconducting states of Nb using a pure Nb sheet. The temperature dependence of thermal conductivity of the Nb sheet was measured under magnetic field using TTO of PPMS. A huge MTSR of 650% was observed at extreme conditions of $T$ = 2.5 K and $H$ = 4.0 kOe. Furthermore, moderate MTSR exceeding 200% was observed in a wide $H$-$T$ range, suggesting flexibility of MTS performance to temperature and magnetic field. This work will open new pathway to explore novel thermal switching for low-temperature electronic devices.

**Acknowledgments**

The authors thank O. Miura for supports in magnetization experiments. The work was partly supported by JST-ERATO (JPMJER2201) and Tokyo Government Advanced Research (H31-1).

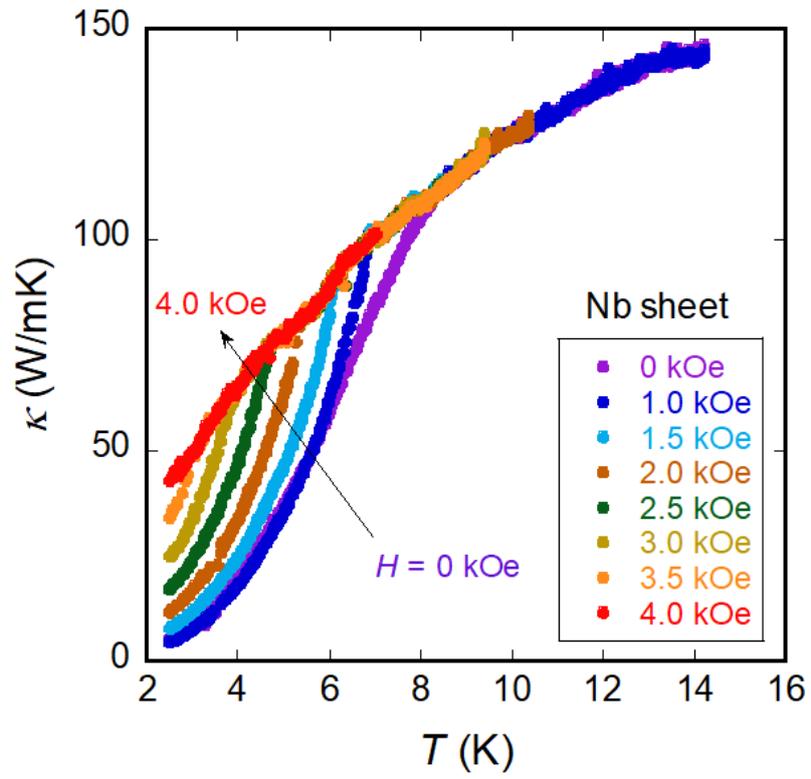

Fig.1. Temperature dependences of $\kappa$ of the Nb sheet under magnetic fields up to 4.0 kOe.

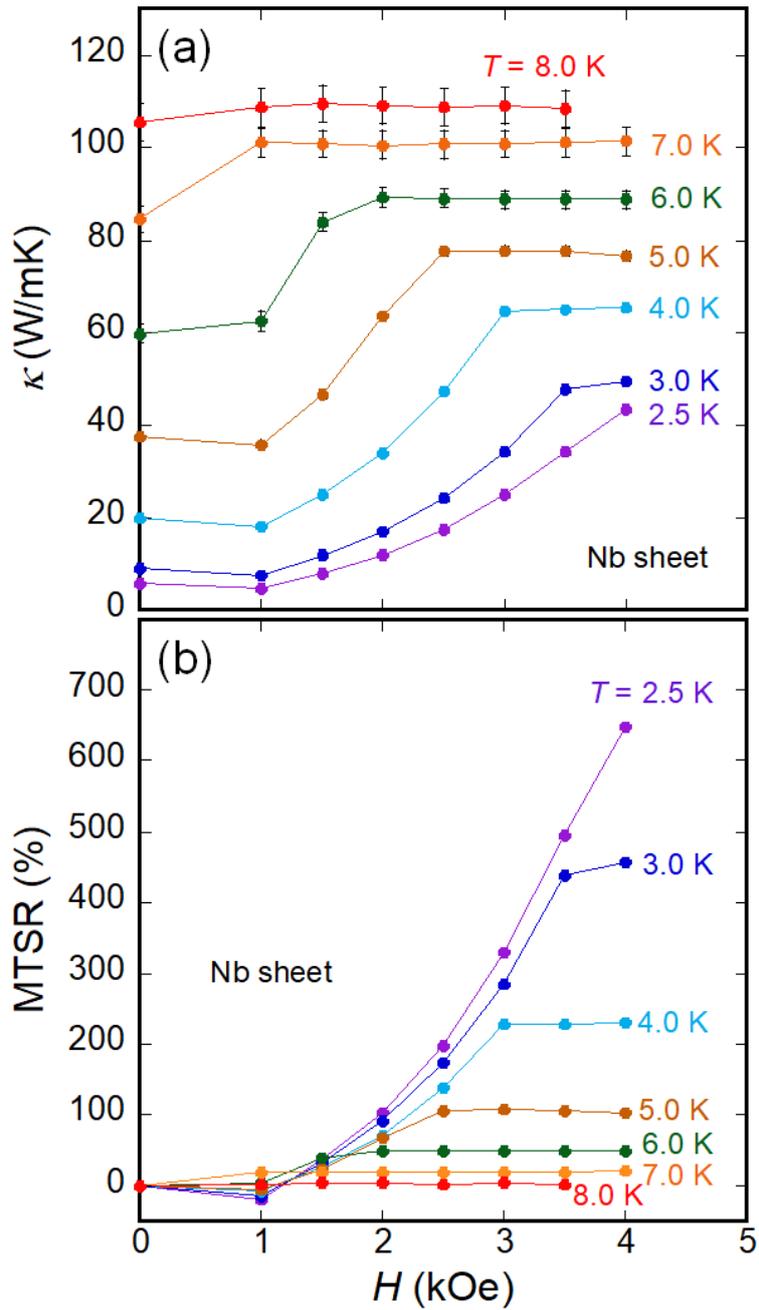

Fig.2. (a) Magnetic field dependences of $\kappa$ of the Nb sheet at various temperatures. (b) Estimated magneto-thermal-switching ratio (MTSR) as a function of magnetic field.

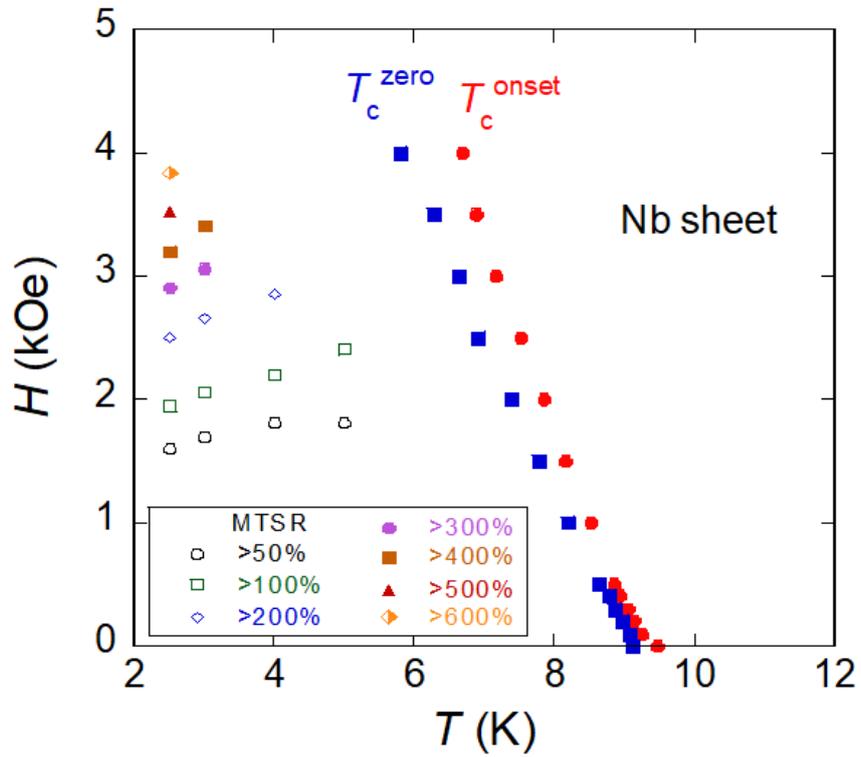

Fig.3. Magnetic field-temperature phase diagram of the superconducting states of Nb. $H_{c2}$ and $H_{irr}$ denote the upper critical field and irreversibility filed estimated from the onset and zero-resistivity temperatures, respectively. The data of MTSR indicates the conditions of magnetic field and temperature at which the MTSR is achieved.

# Supplementary data

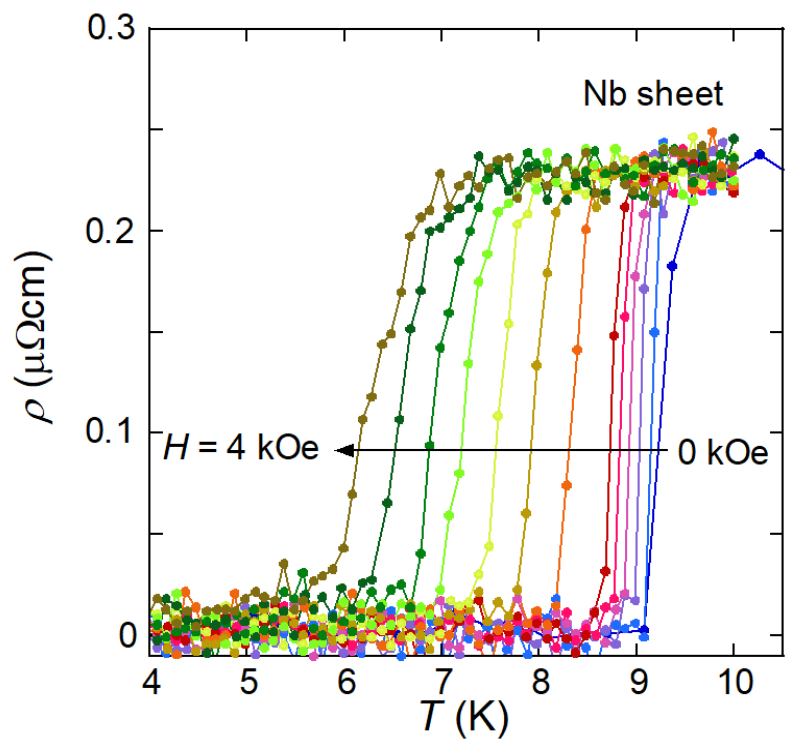

Fig. S1. Temperature dependences of electrical resistivity ($\rho$) of the Nb sheet under magnetic fields up to 4.0 kOe.

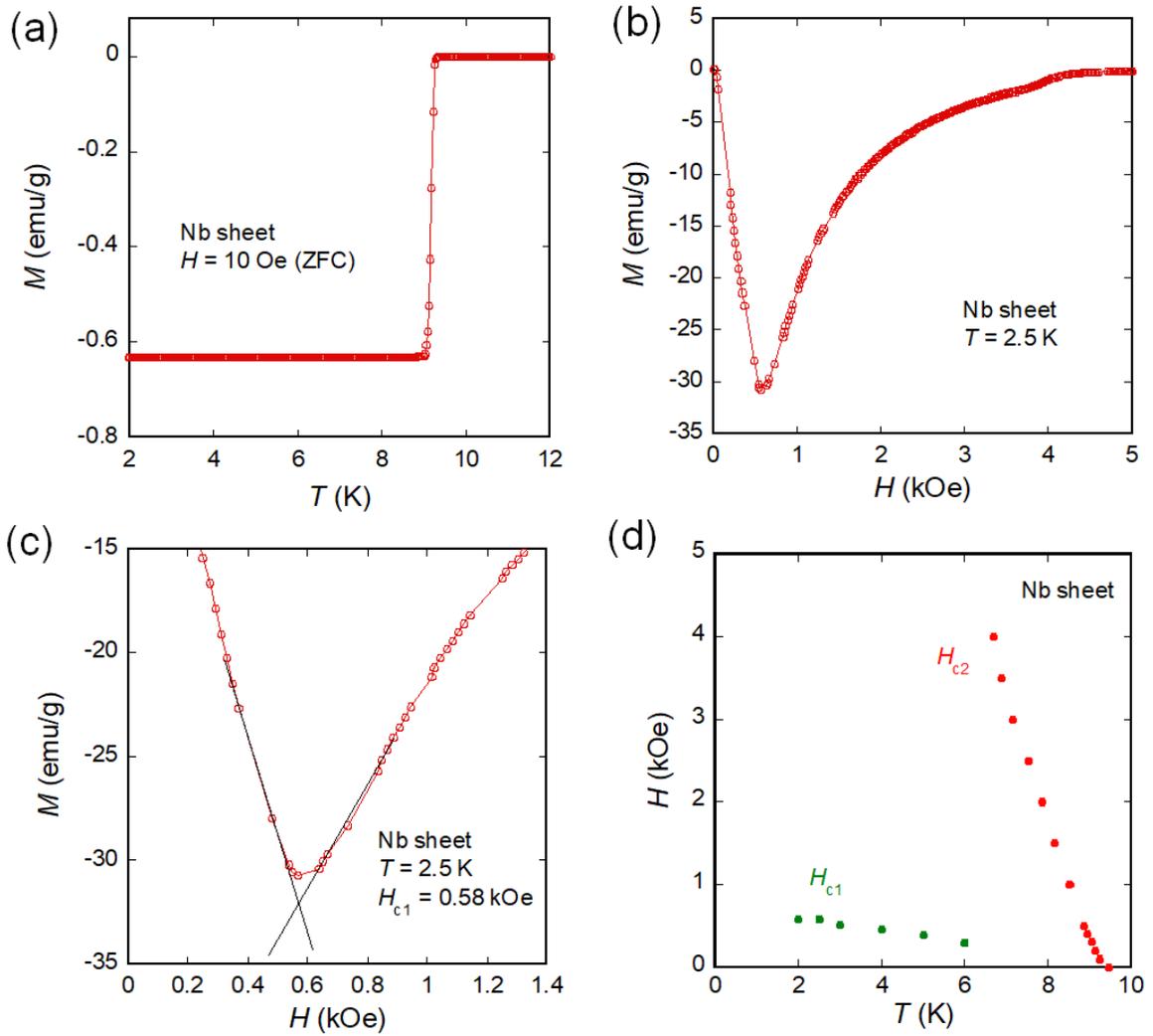

Fig. S2. (a) Temperature dependence of ZFC (zero-field cooling) magnetization of Nb sheet. (b) Magnetic field dependence of magnetization of Nb sheet. (c) Definition of $H_{c1}$ used for the analysis of $M$-$H$ data. (d) Magnetic field-temperature phase diagram. $H_{c1}$ and $H_{c2}$ denote lower critical fields and upper critical fields, respectively.

**Sample #2**

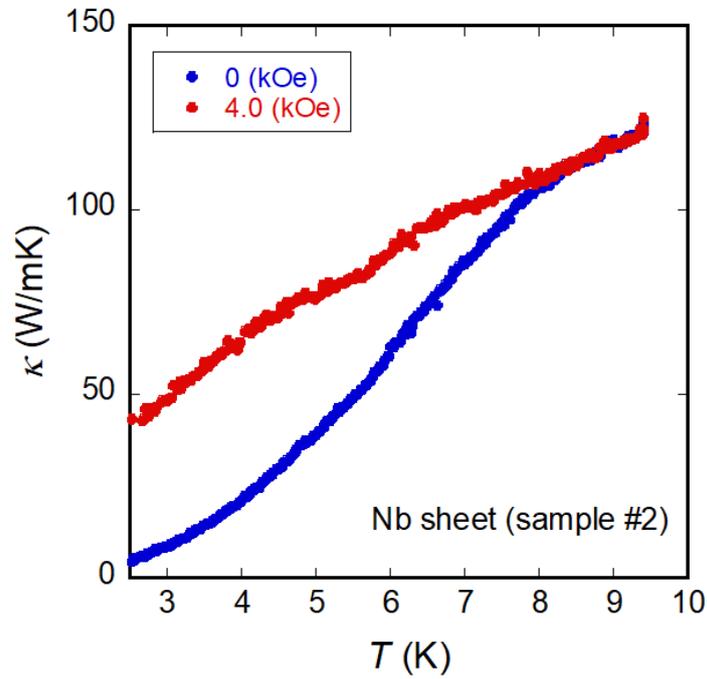

Fig. S3. Temperature dependence of $\kappa$ for the Nb sheet (sample #2) under 0 and 4.0 kOe. The terminals of the Thermal Transport Option unit were directly attached to the Nb sheet.

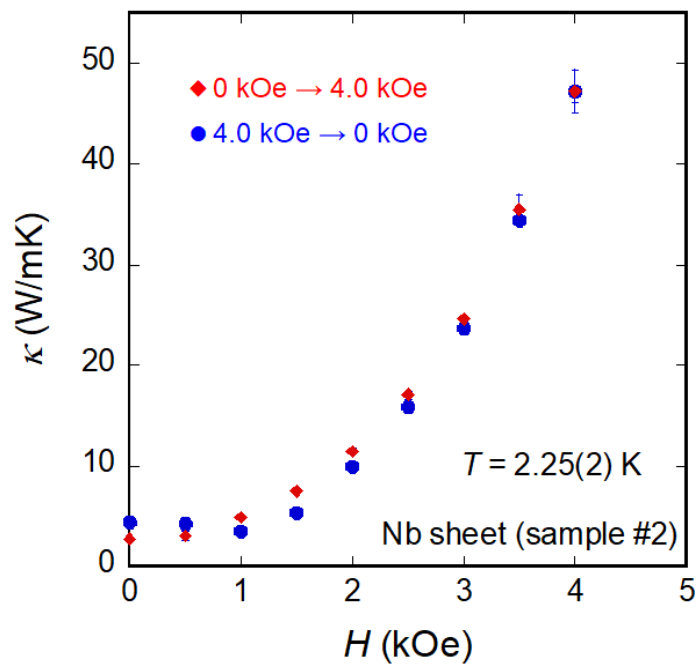

Fig. S4. Magnetic field dependence of $\kappa$ for the Nb sheet (sample #2) at $T = 2.25(2)$ K.